\documentclass
[superscriptaddress,secnumarabic,amssymb,amsmath,nobibnotes,aps,prd,showkeys,showpacs,nofootinbib,onecolumn,12pt]{revtex4}%
\usepackage{graphicx}
\usepackage{epsf}
\usepackage{bm}
\usepackage{amsmath}
\usepackage{amsfonts}
\usepackage{amssymb}%
\providecommand{\U}[1]{\protect\rule{.1in}{.1in}}

\newcommand{\be}{\begin{equation}}
\newcommand{\ee}{\end{equation}}

\newcommand{\mincir}{\raise
-3.truept\hbox{\rlap{\hbox{$\sim$}}\raise4.truept\hbox{$<$}\ }}
\newcommand{\magcir}{\raise
-3.truept\hbox{\rlap{\hbox{$\sim$}}\raise4.truept\hbox{$>$}\ }}

\begin{document}
\title{Classical and quantum cosmological solutions in teleparallel dark energy with
anisotropic background geometry}
\author{Andronikos Paliathanasis}
\email{anpaliat@phys.uoa.gr}
\affiliation{Institute of Systems Science, Durban University of Technology, PO Box 1334,
Durban 4000, South Africa}
\affiliation{Instituto de Ciencias F\'{\i}sicas y Matem\'{a}ticas, Universidad Austral de
Chile, Valdivia 5090000, Chile}

\begin{abstract}
We investigate exact and analytic solutions for the field equations in
teleparallel dark energy model where the physical space is described by the
locally rotational symmetric Bianchi I, Bianchi III and Kantowski-Sachs
geometries. We make use of the property that a point-like Lagrangian exist for
the description of the field equations and variational symmetries are applied
for the construction of invariant functions and conservation laws. The later
are used for the derivation of new analytic solutions for the classical field
equations and exact function forms for the wavefunction in the quantum limit.

\end{abstract}
\date{\today}
\maketitle



\section{Introduction}

\label{sec1}

In teleparallelism the gravitational Action Integral is defined by the
torsion scalar $T~$\cite{cc,Hayashi79} which is constructed by the
antisymmetric Weitzenb\"{o}ck connection \cite{Weitzenb23}. Teleparallel
theory of gravity is equivalent to General Relativity and the presence of
the cosmological constant plays does not affect the equivalence of the two
theories. However, that is not true when other geometric invariants are
introduced into the gravitational Action Integral, or scalar fields which
are nonminimally coupled to gravity. There is a plethora of gravitational
theories inspired by teleparallelism \cite{fer1,fer2,fer3,fer7,f01,mm2}, see
also the recent review \cite{revtel} for more details. On the other hand,
scalar fields play an important role in the description of the various
epochs of the universe. The inflationary era is usually attributed to a
scalar field known as inflaton \cite{guth} which drives the dynamics to
provide expansion of the universe, while the dark energy is assumed to be
described by a scalar field. Moreover, various unified dark energy models
have been also proposed in the literature \cite{Ratra88,bsjd,uni1}.

In this piece of work we are interested in a modified teleparallel theory of
gravity known as scalar-torsion theory or teleparallel dark energy model
\cite{sss1,ss2,ss3,ss4} in an anisotropic background space. The Action
Integral is linear to the torsion scalar $T$, however, a scalar field is
introduced which interacts to the gravity in the Lagrangian. The theory can
be seen as extension of the scalar-tensor models in teleparallelism. While
there are many similarities between the scalar-tensor and scalar-torsion
theories, i.e. they are second-order theories with the same degrees of
freedom, the theories are totally different. For instance while the two
theories can admit duality transformations these provides different physical
properties \cite{ss5}. Recently in \cite{sm1}, the cosmological dynamics
were studied for the scalar-tensor and scalar-torsion theories in a
spatially flat isotropic and homogeneous universe. Various cosmological
applications of the scalar-torsion theory can be found for instance in \cite%
{te1,te3,te5}. {Constraints of the scalar-torsion theory with
cosmological observations in a isotropic universe performed in \cite{cn1}.
It was found that the observations favor a nonminimally coupling in
scalar-torsion theory}

The purpose of this piece of work is to provide for the first time analytic
and exact solutions for the classical and quantum limit of the field
equations in scalar-torsion theory with an anisotropic background space.
Specifically, we consider the physical space to be described by the locally
rotational spacetimes of Bianchi I, Bianchi III and Kantowski-Sachs
spacetimes. These locally rotational spacetimes are homogeneous and
anisotropic with two scale factors and four isometries. In the limit of
isotropization the spacetimes are reduced to the spatially flat, closed and
open Friedmann--Lema\^{\i}tre--Robertson--Walker geometries. Anisotropic
spacetimes are of special interest because they can describe the
pre-inflationary \cite{sm2} era and they can be used as toy models for the
description of the small anisotropies in the observed universe \cite{sm3}.
Kasner and Kasner-like are exact solutions of Bianchi I geometry. In
teleparallelism this kind of solution was investigated before in \cite%
{sm4,sm6}, while some other studies on exact and analytic solutions of a
Bianchi I geometry in teleparallelism can be found for instance in \cite%
{sm7,sm8,sm9}. In a series of studies, the evolution of the dynamics for the
anisotropic parameters investigated in a higher-order modified theory of
gravity \cite{sm10} and references therein. It was found that independently
of the initial conditions of the cosmological model on the anisotropy and on
the spatially space curvature the universe evolves to an isotropic and
spatially flat geometry.

{The study of anisotropic cosmologies is important for a
gravitational theory. According to the cosmological principle in large
scales the universe is assumed to be isotropic and homogeneous. Inflation is
the a mechanism which has been proposed to solved the isotropization of the
universe, however, anisotropies may play an important role in the
pre-inflation epoch. In his famous paper, R. Wald \cite{rw} found that the
existence of a positive cosmological constant in the context of General
Relativity in anisotropic background geometries leads to isotropic universes
in large scales, which lead to the cosmic no-hair conjecture. Additionally,
the anisotropic spacetimes studied in this work are related with important
initial singularities. Indeed, the dynamical variables of Bianchi I
spacetime describe the behaviour of the Mixmaster universe near the
cosmological singularity. On the other hand, Bianchi III and Kantowski-Sachs
spacetimes can also describe the dynamics of the physical parameters to the
interior of a black hole \cite{pan1,pan2,pan3}. While exact and analytic
solutions for isotropic geometries have been bound before for the case of
scalar-torsion theory, according to our knowledge no anisotropic solutions
have been derived in the literature. The study of the integrability
properties for the field equations and the derivation of solutions is
essential, because we can infer that actual solutions exist for the
description of the physical parameters in scalar-torsion theory with
anisotropic background geometry. }

The gravitational field equations for the model of our analysis are
nonlinear second-order differential equations.\ Hence, mathematical
techniques which deal with nonlinear dynamical systems should be enrolled in
order to determine exact and analytic solutions.\ The field equations for
this specific cosmological model have the property to be described by the
variation of a point-like Lagrangian functions. Hence, in order to construct
conservation laws and invariant functions which are to be used to the
derivation of solutions, Noether's theorem for point transformations is
considered \cite{ns20}. The main concept of the Noether symmetry analysis is
to constrain the unknown functions and parameters of the gravitational
theory, such that the Action Integral is invariant under the application of
point transformations, where according to Noether's second theorem
conservation laws exist \cite{ns01}. Thus it is feasible to infer about the
integrability properties of the field equations and to determine analytic
solutions. Furthermore, the existence of the point-like Lagrangian means
that the field equations can be written also by using the Hamiltonian
formalism. Thus, under the classical quantization process, the
Wheeler-DeWitt equation of quantum cosmology can be written \cite{wd1}. The
Noetherian conservation laws for the classical field equations provide
differential invariants for the Wheeler-DeWitt equations, which are
necessary in order to determine closed-form expression for the wavefunction
as described by the Wheeler-DeWitt equation \cite{wdw2}. For a review on the
Wheeler-DeWitt equation we refer the reader to \cite{wdw6} The structure of
this paper is as follows.

In Section \ref{sec2} we briefly present the basic definitions of
teleparallelism and we define the cosmological model of this study, which is
that of teleparallel dark energy. Anisotropic background spacetimes are
considered and the field equations are derived for physical spaces described
by the locally rotational Bianchi I, Bianchi III and Kantowski-Sachs
geometries. In Section \ref{sec4} we apply the conditions provided by
Noether's first theorem in order to constrain the unknown functional form of
the scalar field potential where Noether symmetries exist. These results are
applied in order to infer the Liouville integrability of the field
equations. Classical exact and analytic solutions of the field equations are
constructed in Section \ref{sec5}. Furthermore, in Section \ref{sec6} we
write the Wheeler-DeWitt equation and with the application of the
differential invariants provided by the classical conservation laws we
determine exact solutions for the wavefunction. Finally, in Section \ref%
{sec7} we summarize our results.

\section{Teleparallel dark energy}

\label{sec2}

In teleparallelism the fundamental geometric objects are the vierbein fields
$e_{i}$ defined by the constraint $g(e_{i},e_{j})=e_{i}.e_{j}=\eta_{ij}~$%
\cite{cc}, in which $\eta_{ij}=\mathrm{diag}(-1,1,1,1)$ is the Lorentz
metric in canonical form.

In terms of coordinates with a nonholonomic basis $e^{i}(x^{\kappa})=h_{\mu
}^{i}(x^{\kappa})dx^{i}$, the metric tensor $g_{\mu\nu}(x^{\kappa})$ the
metric tensor is%
\begin{equation}
g_{\mu\nu}=\eta_{ij}h_{~\mu}^{i}h_{~\nu}^{j}.  \label{d.04}
\end{equation}

We define the teleparallel torsion tensor
\begin{equation}
T_{\mu \nu }^{~~\beta }=\hat{\Gamma}_{\nu \mu }^{\beta }-\hat{\Gamma}_{\mu
\nu }^{\beta }=h_{~~i}^{\beta }(\partial _{\mu }h_{~\nu }^{i}-\partial _{\nu
}h_{~\mu }^{i})~,  \label{d.05}
\end{equation}%
which is the antisymmetric part of the affine connection coefficients \cite%
{Weitzenb23}.

In General Relativity the gravitational Lagrangian is defined by the
Ricciscalar of the Levi-Civita. In teleparallelism the gravitational
Lagrangian is defined by the torsion scalar $T$ related with the torsion
tensor $T_{\mu\nu}^{~~\beta}$.

Specifically the scalar $T$ is defined as
\begin{equation}
T=S_{~\beta}^{\mu\nu}T_{~\mu\nu}^{\beta}
\end{equation}
in which $S_{~\beta}^{\mu\nu}=\frac{1}{2}(K_{~\beta}^{\mu\nu}+\delta_{~\beta
}^{\mu}T_{~\theta}^{\theta\nu}-\delta_{\beta}^{\nu}T_{~\theta}^{\theta\mu})$
and $K_{~\beta}^{\mu\nu}=-\frac{1}{2}(T_{~\beta}^{\mu\nu}-T_{~\beta}^{\nu\mu
}-T_{~\beta}^{\mu\nu})$~is the contorsion tensor which equals the difference
of the Levi Civita connection in the holonomic and the nonholonomic frame.

The gravitational action integral in teleparallel equivalence of General
Relativity is defined as \cite{cc}
\begin{equation}
S=\frac{1}{16\pi G}\int d^{4}xe\left( T+L_{m}\right) ,
\end{equation}%
where $L_{m}$ is the Lagrangian component for the matter source and $e=\sqrt{%
-g}$.

Similar to the case of General Relativity, scalar fields have been
introduced in teleparallelism. Inspired by the Brans-Dicke theory, i.e. the
scalar-tensor theories of General Relativity, a Machian teleparallel
gravitational theory has been considered. It is called teleparallel dark
energy, or scalar-tensor teleparallel gravity and it is defined by the
gravitational Action Integral \cite{sss1,ss2,ss3,ss4,ss5}%
\begin{equation}
S=\frac{1}{16\pi G}\int d^{4}xe\left[ F\left( \phi\right) \left( T+\frac{%
\omega}{2}\phi_{;\mu}\phi^{\mu}+V\left( \phi\right) \right) \right] ,
\label{d.08}
\end{equation}
where $F\left( \phi\right) $ is the coupling function of the scalar field
with the torsion scalar, $\omega$ is a constant nonzero parameter, analogue
of the Brans-Dicke parameter, and $V\left( \phi\right) $ is the scalar field
potential which drives the dynamics.

An equivalent way to write the Action Integral (\ref{d.08}) is
\begin{equation}
S=\frac{1}{16\pi G}\int d^{4}xe\left[ \hat{F}\left( \psi\right) T+\frac {1}{2%
}\psi_{;\mu}\psi^{\mu}+\hat{V}\left( \psi\right) \right] ,  \label{d.08a}
\end{equation}
where now the new field $\psi$ is related with $\phi$ as $d\psi=\sqrt{\omega
F\left( \phi\right) }d\phi$. Note that, when $\hat{F}\left( \psi\right)
=F_{0}\psi^{2}$ or $F\left( \phi\right) =F_{0}e^{2\phi}$, the given
gravitational model is the analogue of the Brans-Dicke theory, or of the
dilaton equivalent model in teleparallelism.

The teleparallel dilaton model is of special interest because it admits
similar properties with the usual dilaton theory. Indeed, in the case of a
spatially flat Friedmann--Lema\^{\i}tre--Robertson--Walker (FLRW) background
geometry it has been shown that the field equations admit a discrete
symmetry which is analogue to the duality Gasperini-Veneziano transformation
\cite{ss5}.

\subsection{Anisotropic spacetimes}

In the following analysis we consider the physical space to be homogeneous
and anisotropic described by the generic line element
\begin{equation}
ds^{2}=-N^{2}\left( t\right) dt^{2}+e^{2\alpha\left( t\right) }\left(
e^{2\beta\left( t\right) }dx^{2}+e^{-\beta\left( t\right) }\left(
dy^{2}+f^{2}\left( y\right) dz^{2}\right) \right)  \label{ch.03}
\end{equation}
in which the function $f\left( y\right) $ has one of the following
functional forms $f_{A}\left( y\right) =1$, $f_{B}\left( y\right)
=\sinh\left( y\right) $ or $f_{C}\left( y\right) =\sin\left( y\right) $ such
that the spacetime to be locally rotational, $\alpha\left( t\right)
,~\beta\left( t\right) $ are the two free scale factors and $N\left(
t\right) $ is the lapse function. For $f\left( y\right) =f_{A}\left(
y\right) $, the line element (\ref{ch.03}) is that of the Bianchi\ I space,
for $f\left( y\right) =f_{B}\left( y\right) $ the Bianchi III space is
recovered while, when $f\left( y\right) =f_{C}\left( y\right) $, the line
element (\ref{ch.03}) is that of the Kantowski-Sachs universe. The parameter
$\alpha\left( t\right) $ is the expansion rate of the three-dimensional
hypersurfaces and $\beta\left( t\right) $ is the anisotropic parameter. When
$\beta\left( t\right) =const$., the line element (\ref{ch.03}) \ describes
an isotropic physical space, that is, the flat, open and closed FLRW
geometries are recovered.

In order to calculate the torsion scalar we should define the proper
vierbein fields. In order for the limit of General Relativity to be
recovered the vierbein fields should be defined properly. We follow the
discussion in \cite{revtel} for the Bianchi I spacetime we assume the
vierbein basis
\begin{equation*}
e^{1}=Ndt~,~e^{2}=e^{\alpha+\beta}dx~,~e^{3}=e^{\alpha-\frac{\beta}{2}%
}dy~,~e^{4}=e^{\alpha-\frac{\beta}{2}}dz
\end{equation*}
with torsion scalar%
\begin{equation}
T_{B_{I}}=\frac{1}{N^{2}}\left( 6\dot{\alpha}^{2}-\frac{3}{2}\dot{\beta}%
^{2}\right) .  \label{ch.04}
\end{equation}

For Bianchi III we consider the basis%
\begin{align*}
e^{1}& =Ndt~, \\
e^{2}& =i~e^{\alpha +\beta }\cos z\sinh y~dx+e^{\alpha -\frac{\beta }{2}%
}\left( \cosh y\cos z~dy-\sinh y\sin z~dz\right) ~, \\
e^{3}& =i~e^{\alpha +\beta }\sinh y\sin z~dx+e^{\alpha -\frac{\beta }{2}%
}\left( \cosh y\sin z~dy-\sinh y\cos z~dz\right) ~, \\
e^{4}& =-e^{\alpha +\beta }\cosh y~dx-i~e^{\alpha -\frac{\beta }{2}}\sinh
y~dy~,
\end{align*}%
in which we calculate the torsion scalar
\begin{equation}
T_{B_{III}}=\frac{1}{N^{2}}\left( 6\dot{\alpha}^{2}-\frac{3}{2}\dot{\beta}%
^{2}\right) +2e^{-2\alpha +\beta }.  \label{ch.05}
\end{equation}

Finally, for the Kantowski-Sachs spacetime we assume the vierbein fields%
\begin{align*}
e^{1}& =Ndt~, \\
e^{2}& =e^{a+\beta }\cos z\sin y~dx+e^{a-\frac{\beta }{2}}\left( \cos y\cos
z~dy-\sin y\sin z~dz\right) ~, \\
e^{3}& =e^{a+\beta }\sin y\sin z~dx+e^{a-\frac{\beta }{2}}\left( \cos y\sin
z~dy-\sin y\cos z~dz\right) ~, \\
e^{4}& =e^{a+\beta }\cos y~dx-e^{a-\frac{\beta }{2}}\sin y~dy,
\end{align*}%
from which we calculate the torsion scalar

\begin{equation}
T_{KS}=\frac{1}{N^{2}}\left( 6\dot{\alpha}^{2}-\frac{3}{2}\dot{\beta}%
^{2}\right) -2e^{-2\alpha+\beta}.  \label{ch.06}
\end{equation}

Thus, we define the general torsion scalar
\begin{equation*}
T_{A}\left( K\right) =\frac{1}{N^{2}}\left( 6\dot{\alpha}^{2}-\frac{3}{2}%
\dot{\beta}^{2}\right) +2Ke^{-2\alpha+\beta},
\end{equation*}
where $K=0$ is for the Bianchi I space, $K=1$ is for the Bianchi I space and
$K=-1$ corresponds to the torsion scalar of the Kantowski-Sachs space.

By replacing the generic $T_{A}\left( K\right) $ in the Action Integral (\ref%
{d.08}) and assuming that the scalar field inherits the isometries of the
background space, that is, $\phi=\phi\left( t\right) $, we derive the
point-like Lagrangian for the field equations%
\begin{equation}
L\left( N,a,\dot{a},\beta,\dot{\beta},\phi,\dot{\phi}\right) =F\left(
\phi\right) e^{3\alpha}\left( \frac{1}{N}\left( 6\dot{\alpha}^{2}-\frac {3}{2%
}\dot{\beta}^{2}-\frac{\omega}{2}\dot{\phi}^{2}\right) +N\left(
2Ke^{-2\alpha+\beta}+V\left( \phi\right) \right) \right)  \label{ch.07}
\end{equation}
while the field equations are
\begin{equation}
0=F\left( \phi\right) e^{3\alpha}\left( 6\dot{\alpha}^{2}-\frac{3}{2}\dot{%
\beta}^{2}-\frac{\omega}{2}\dot{\phi}^{2}-V\left( \phi\right)
-2Ke^{-2\alpha+\beta}\right) ~,  \label{ch.08}
\end{equation}%
\begin{equation}
0=\ddot{\alpha}+\frac{3}{2}\dot{\alpha}^{2}+\frac{3}{8}\dot{\beta}^{2}+\frac{%
1}{4}\left( \frac{\omega}{2}\dot{\phi}^{2}-V\left( \phi\right) \right) +%
\frac{d}{dt}\left( \ln F\left( \phi\right) \right) \dot{\alpha }-\frac{1}{6}%
Ke^{-2\alpha+\beta}~,  \label{ch.09}
\end{equation}%
\begin{equation}
0=\ddot{\beta}+3\dot{\alpha}\dot{\beta}+\frac{d}{dt}\left( \ln F\left(
\phi\right) \right) \dot{\beta}+\frac{2}{3}Ke^{-2\alpha+\beta}~,
\label{ch.10}
\end{equation}%
\begin{equation}
0=\omega\left( \ddot{\phi}+3\dot{\alpha}\dot{\phi}\right) \dot{\phi}+\dot {V}%
+\frac{d}{dt}\left( \ln F\left( \phi\right) \right) \left( 6\dot {a}^{2}-%
\frac{3}{2}\dot{\beta}^{2}+V\left( \phi\right) +2Ke^{-2\alpha+\beta }\right)
~,  \label{ch.11}
\end{equation}
where without loss of generality we have selected the lapse function $%
N\left( t\right) $ to be constant, i.e. $N\left( t\right) =1$.

\section{Symmetry analysis}

\label{sec4}

In this section, we proceed with the derivation of variational symmetries
for the field equations. Because the field equations for the cosmological
model of our consideration follow from a point-like Lagrangian the
variational symmetries are derived with the application of Noether's
theorem. In this scenario variational symmetries are called Noether
symmetries, for a recent review on Noether's work we refer the reader in
\cite{ns01}.

Noether's theorem in modified theories of gravity cover a wide range of
applications with many important results on the classification for the free
functions of the modified theories and the construction of conservation laws
which have been used for the derivation of analytic and exact solutions \cite%
{ns02,ns03,ns04,ns07,ns15,ns19,ns20}. For convenience of the reader we
discuss the main definitions of Noether's theorems.

\subsection{Noether's theorems}

Consider the action integral
\begin{equation}
A=\int L\left( t,q\left( t\right) ,\dot{q}\left( t\right) \right) dt.
\label{4421}
\end{equation}%
where $L\left( t,q\left( t\right) ,\dot{q}\left( t\right) \right) $ is the
Lagrangian function, $t$ is the independent variable and $q\left( t\right) $
denotes the dependent variable while $\dot{q}\left( t\right) =\frac{dq\left(
t\right) }{dt}$ is the first-order derivative

Then, under the infinitesimal transformation at any point $P$ in the space
of variables $\left\{ t,q\right\} $%
\begin{equation}
\bar{t}=t+\varepsilon \xi \left( t,q\right) ,\qquad \bar{q}=q+\varepsilon
\eta \left( t,q\right)  \label{4422}
\end{equation}%
with infinitesimal generator the vector field
\begin{equation}
X=\xi \left( t,q\right) \partial _{t}+\eta \left( t,q\right) \partial _{q},
\end{equation}%
the action integral (\ref{4421}) becomes
\begin{equation}
\bar{A}=\int_{\bar{t}_{0}}^{\bar{t}_{1}}L\left( \bar{t},\bar{q},\dot{\bar{q}}%
\right) d\bar{t}.
\end{equation}

Thus, according to Noether's first theorem, the variation of the action
integral $A$ remains invariant under the application of the infinitesimal
transformation (\ref{4422}) if and only if there exists a function $f$ such
that \cite{ns20}
\begin{equation}
\dot{f}=\xi \frac{\partial L}{\partial t}+\eta \frac{\partial L}{\partial q}%
+\left( \dot{\eta}-\dot{q}\dot{\xi}\right) \frac{\partial L}{\partial \dot{q}%
}+\dot{\xi}L~.  \label{ff3}
\end{equation}%
The function $f$ is a boundary term which has been introduced to allow for
the infinitesimal changes in the value of the action integral provided by
the infinitesimal transformation.

The novelty of Noether's work is that there exists a simple formula for the
one-to-one correspondence between the symmetry vectors and conservation laws
for the equations of motion. Indeed, if $X$ satisfies condition (\ref{ff3})
for the Lagrangian function $L\left( t,q\left( t\right) ,\dot{q}\left(
t\right) \right) $, then function \cite{ns20}%
\begin{equation}
I\left( X\right) =f-\left[ \xi L+\left( \eta -\dot{q}\xi \right) \frac{%
\partial L}{\partial \dot{q}}\right]  \label{ff4}
\end{equation}%
is a conservation law for the equations of motion; that is, $\dot{I}\left(
X\right) =0$. Formula (\ref{ff4}) is the so-called Noether's second theorem.

\subsection{Symmetry classification}

In this study and for the point-like Lagrangian (\ref{ch.07}) with $F\left(
\phi \right) =e^{2\phi },$ where without loss of generality, we assume $%
N\left( t\right) =e^{3\alpha +2\phi }$, that is,%
\begin{equation}
L\left( N,a,\dot{a},\beta ,\dot{\beta},\phi ,\dot{\phi}\right) =\left( 6\dot{%
\alpha}^{2}-\frac{3}{2}\dot{\beta}^{2}-\frac{\omega }{2}\dot{\phi}%
^{2}\right) +e^{6\alpha +4\phi }\left( 2Ke^{-2\alpha +\beta }+V\left( \phi
\right) \right) ,  \label{ff5}
\end{equation}%
we consider the infinitesimal transformation
\begin{align}
\bar{t}& =t+\varepsilon \xi \left( t,\alpha ,\beta ,\phi \right) ~, \\
\bar{\alpha}& =\alpha +\varepsilon \eta ^{\alpha }\left( t,\alpha ,\beta
,\phi \right) ~, \\
\bar{\beta}& =\beta +\varepsilon \eta ^{\beta }\left( t,\alpha ,\beta ,\phi
\right) ~, \\
\bar{\phi}& =\phi +\varepsilon \eta ^{\phi }\left( t,\alpha ,\beta ,\phi
\right)
\end{align}%
with generator the vector field $X=\xi \partial _{t}+\eta ^{\alpha }\partial
_{\alpha }+\eta ^{\beta }\partial _{\beta }+\eta ^{\phi }\partial _{\phi }$.

Hence, application of the symmetry condition (\ref{ff3}) for the Lagrangian
function (\ref{ff5}) gives a system of linear partial differential equations
which determine the generator $X$ for various functional forms of the
potential $V\left( \phi\right) $ and values of the curvature term $K$.

For the Bianchi I background space, $K=0$, it follows that, for a zero
potential function $V\left( \phi \right) =0$, the admitted Noether
symmetries by the field equations are
\begin{equation}
X_{1}=\partial _{\alpha }~,~X_{2}=\phi \partial _{\alpha }+\frac{12}{\omega }%
\alpha \partial _{\phi }~,~X_{3}=\beta \partial _{\alpha }+4\alpha \partial
_{\beta }~,
\end{equation}%
\begin{equation}
X_{4}=\frac{\omega }{3}\phi \partial _{\beta }-\beta \partial _{\phi
}~,~X_{5}=\partial _{\beta }~,~X_{6}=\partial _{\phi }~.
\end{equation}%
The symmetry vectors are the isometries of the three-dimensional flat space,
$\left\{ X_{1},X_{5},X_{6}\right\} $ are the translation symmetries while $%
\left\{ X_{2},X_{3},X_{4}\right\} $ are the three rotations of the
three-dimensional minisuperspace.

The corresponding conservation laws are calculated from expression (\ref{ff4}%
) and they are
\begin{equation}
I\left( X_{1}\right) =\dot{\alpha}~,~I\left( X_{2}\right) =\phi \dot{\alpha}%
+\alpha \dot{\phi}~,~I\left( X_{3}\right) =\beta \dot{\alpha}+\alpha \dot{%
\beta}
\end{equation}
\begin{equation}
I\left( X_{4}\right) =\phi \dot{\beta}-\beta \dot{\phi}~,~I\left(
X_{6}\right) =\dot{\beta}~\text{and }I\left( X_{6}\right) =\dot{\phi}~\text{.%
}
\end{equation}

On the other hand, for $K=0$ and $V\left( \phi \right) =V_{0}e^{\lambda \phi
}$, the additional admitted Noether symmetries from the Lagrangian of the
field equations are
\begin{equation}
X_{5}~,~X_{7}=\frac{\left( 4+\lambda \right) }{2\omega }X_{3}+\frac{3}{%
\omega }X_{4}~,~X_{8}=\left( 4+\lambda \right) X_{1}-X_{6}~,
\end{equation}%
for which the resulting conservation laws are
\begin{equation}
I\left( X_{5}\right) ~,~I\left( X_{8}\right) =\left( 4+\lambda \right) \dot{a%
}+\omega \dot{\phi}~,
\end{equation}%
\begin{equation}
I\left( X_{7}\right) =2\left( 4+\lambda \right) B\dot{\alpha}-\left( 2\alpha
\left( 4+\lambda \right) +\omega \phi \right) \dot{\beta}+\omega \beta \dot{%
\phi}~.
\end{equation}

The admitted Noether symmetries are different in the presence of the
curvature term, $K\neq 0$. In this case, for the zero potential $V\left(
\phi \right) =0$, the admitted\ Noether symmetries are
\begin{equation}
Y_{1}=X_{1}-4X_{5}~,~Y_{2}=X_{6}-4X_{5}~,
\end{equation}%
and%
\begin{equation}
Y_{3}=X_{2}-\frac{12}{\omega }X_{4}~,
\end{equation}%
where the resulting conservation laws are%
\begin{equation}
I\left( Y_{1}\right) =\dot{\alpha}+\dot{\beta}~,~I\left( Y_{2}\right) =12%
\dot{\beta}-\omega \dot{\phi}~,
\end{equation}%
\begin{equation}
I\left( Y_{3}\right) =\phi \left( \dot{\alpha}+\dot{\beta}\right) -\left(
\alpha -\beta \right) \dot{\phi}\text{.}
\end{equation}

Finally, for the exponential potential $V\left( \phi\right)
=V_{0}e^{\lambda\phi}$ the field equations admit the unique symmetry vector
\begin{equation}
Y_{4}=\left( \lambda+4\right) X_{1}-4\left( \lambda-2\right) X_{5}-6X_{6}~
\end{equation}
with conservation law%
\begin{equation}
I\left( Y_{4}\right) =\left( \lambda+4\right) \dot{\alpha}+\left(
\lambda-2\right) \dot{\beta}-\frac{\omega}{2}\dot{\phi}~.
\end{equation}

From the symmetry analysis, we conclude that the field equations form a
Liouville integrable dynamical system according to the Noether point
symmetries for the case of Bianchi I for $V\left( \phi \right) =0$ and $%
V\left( \phi \right) =V_{0}e^{\lambda \phi }$, and for the Bianchi III and
Kantowski-Sachs cases $K\neq 0$ only when $V\left( \phi \right) =0$.

\section{Classical solutions}

\label{sec5}

We proceed with the application of the conservation laws for the derivation
of analytic solutions for the latter cases.

\subsection{Bianchi I spacetime}

For the case of Bianchi I spacetime, for $4+\lambda \neq 0~$we define the
new dependent variable $\Phi =\phi +\frac{6}{4+\lambda }\alpha $ such that
the point-like Lagrangian (\ref{ff5}) reads%
\begin{equation}
L\left( \alpha ,\dot{\alpha},\beta ,\dot{\beta},\Phi ,\dot{\Phi}\right)
=\left( 6-\frac{18\omega }{\left( 4+\lambda \right) ^{2}}\right) \dot{\alpha}%
^{3}-\frac{3}{2}\dot{\beta}^{2}+\frac{6\omega }{4+\lambda }\dot{\alpha}\dot{%
\Phi}-\frac{1}{2}\omega \dot{\Phi}^{2}+V_{0}e^{\left( 4+\lambda \right) \Phi
}.
\end{equation}

This gives the field equations
\begin{equation}
\left( 6-\frac{18\omega }{\left( 4+\lambda \right) ^{2}}\right) \dot{\alpha}%
^{2}-\frac{3}{2}\dot{\beta}^{2}+\frac{6\omega }{4+\lambda }\dot{\alpha}\dot{%
\Phi}-\frac{1}{2}\omega \dot{\Phi}^{2}-V_{0}e^{\left( 4+\lambda \right) \Phi
}=0~,
\end{equation}%
\begin{equation}
\ddot{\alpha}-\frac{V_{0}}{2}e^{\left( 4+\lambda \right) \Phi }=0~,~\ddot{%
\beta}=0~,
\end{equation}%
and%
\begin{equation}
\ddot{\Phi}+\frac{V_{0}\left( \left( 4+\lambda \right) ^{2}-3\omega \right)
}{\left( 4+\lambda \right) \omega }e^{\left( 4+\lambda \right) \Phi }=0.
\end{equation}

Thus, for the anisotropic parameter $\beta $ we derive the closed-form
expression $\beta \left( t\right) =\beta _{1}t+\beta _{0}$.

\subsubsection{Arbitrary parameters}

For $\frac{V_{0}\left( \left( 4+\lambda \right) ^{2}-3\omega \right) }{%
\left( 4+\lambda \right) \omega }\neq 0,$ the analytic solution is
\begin{equation}
\Phi \left( t\right) =-\frac{1}{4+\lambda }\ln \left( \frac{2V_{0}\left(
\left( 4+\lambda \right) ^{2}-3\omega \right) }{\left( 4+\lambda \right)
^{2}\omega \Phi _{1}}\cosh ^{2}\left( \frac{4+\lambda }{2}\sqrt{\Phi _{1}}%
\left( t-t_{0}\right) \right) \right)
\end{equation}%
and
\begin{equation}
\alpha \left( t\right) =\frac{\omega }{\left( \left( 4+\lambda \right)
^{2}-3\omega \right) }\ln \left( \cosh \left( \frac{4+\lambda }{2}\sqrt{\Phi
_{1}}\left( t-t_{0}\right) \right) \right) +\alpha _{1}t+\alpha _{0}
\end{equation}%
with constraint equation
\begin{equation}
\frac{12\alpha _{1}^{2}\left( \left( 4+\lambda \right) ^{2}-3\omega \right)
^{2}-\omega \left( \lambda +4\right) \Phi _{1}}{2\left( \lambda +4\right)
^{4}-\left( \lambda +4\right) ^{2}\omega }-\frac{3}{2}\beta _{1}^{2}=0.
\end{equation}

Consider now the special solution with $\alpha _{1}=0$,~$a_{0}=0,$ and
without loss of generality $t_{0}=0$. This is an anisotropic solution
because from the latter constraint it follows $\beta _{1}\neq 0.\ $That is,
we consider the scale factor%
\begin{equation}
\alpha \left( t\right) =\frac{\omega }{\left( \left( 4+\lambda \right)
^{2}-3\omega \right) }\ln \left( \cosh \left( \frac{4+\lambda }{2}\sqrt{\Phi
_{1}}t\right) \right)
\end{equation}%
and the lapse function%
\begin{equation}
N\left( t\right) =\left( \frac{\omega \left( 4+\lambda \right) ^{2}}{%
2V_{0}\left( \left( 4+\lambda \right) ^{2}-3\omega \right) }\Phi _{1}\right)
^{\frac{2}{4+\lambda }}\left( \cosh \left( \frac{4+\lambda }{2}\sqrt{\Phi
_{1}}t\right) \right) ^{\frac{3\omega -4\left( \lambda +4\right) }{\left(
4+\lambda \right) ^{2}-3\omega }}.
\end{equation}

Thus, the expansion rate $\theta \left( t\right) =\frac{1}{3N}\dot{\alpha}$
is determined to be%
\begin{equation}
\left( \theta \left( a\right) \right) ^{2}\simeq 1-a^{-6+\frac{2}{\omega }%
\left( 4+\lambda \right) ^{2}}~,~\alpha =\ln a.
\end{equation}%
The anisotropic parameter $\sigma =\frac{1}{N}\dot{\beta}\left( t\right) $ is%
\begin{equation}
\sigma \left( a\right) \simeq \alpha ^{-3+\frac{4}{\omega }\left( 4+\lambda
\right) }\text{~,~}\alpha =\ln a.
\end{equation}

Thus, for large values of $\alpha $, the anisotropic parameter vanishes,
that is, $\sigma \left( \alpha \right) \rightarrow 0$, for $-3+\frac{4}{%
\omega }\left( 4+\lambda \right) <0$.

\subsubsection{Case $V_{0}\left( \left( 4+\protect\lambda\right) ^{2}-3%
\protect\omega\right) =0$}

Furthermore, in the case for which $V_{0}\left( \left( 4+\lambda \right)
^{2}-3\omega \right) =0$ the analytic solution is different. Indeed, for $%
V_{0}=0$ we derive the closed-form solution
\begin{align}
\alpha \left( t\right) & =\alpha _{1}t+\alpha _{0}, \\
\Phi \left( t\right) & =\Phi _{1}t+\Phi _{0}
\end{align}%
with constraint equation%
\begin{equation}
~\left( 6-\frac{18\omega }{\left( 4+\lambda \right) ^{2}}\right) \alpha
_{1}^{2}-\frac{3}{2}\beta _{1}^{2}+\frac{6\omega }{4+\lambda }\alpha
_{1}\Phi _{1}-\frac{1}{2}\omega \Phi _{1}^{2}=0.
\end{equation}%
For the latter solution and for $\alpha _{0}=0,~\Phi _{0}$, we derive the
expansion rate and the anisotropic parameters to be%
\begin{equation}
\theta ^{2}\left( a\right) \simeq a^{-\frac{3\lambda }{4+\lambda }+\frac{%
2\Phi _{1}}{\alpha _{1}}}~,\sigma ^{2}\left( a\right) \simeq a^{-\frac{%
3\lambda }{4+\lambda }+\frac{2\Phi _{1}}{\alpha _{1}}}~,~\alpha =\ln a~
\end{equation}%
from which we infer that the universe become isotropic for $-\frac{3\lambda
}{4+\lambda }+\frac{2\Phi _{1}}{\alpha _{1}}<0$.

On the other hand, for $\left( \left( 4+\lambda\right) ^{2}-3\omega\right)
=0 $ the analytic solution is%
\begin{equation}
\Phi\left( t\right) =\Phi_{1}t+\Phi_{0},
\end{equation}%
\begin{equation}
\alpha\left( t\right) =\frac{e^{\left( 4+\lambda\right) \left( \Phi
_{0}+t\Phi_{1}\right) }}{2\left( 4+\lambda\right) ^{2}\Phi_{1}^{2}}%
V_{0}+\alpha_{1}t+a_{0}~,
\end{equation}
with constraint equation $-\frac{1}{6}\left( 4+\lambda\right) \Phi
_{1}\left( \left( 4+\lambda\right) \Phi_{1}-12\alpha_{1}\right) -\frac {3}{2}%
\beta_{1}^{2}=0$.

Assume now $\Phi _{0}=0,~\alpha _{1}=0$ and $\Phi _{1}=0$. In this case the
expansion rate and the anisotropic parameter are%
\begin{equation}
\theta ^{2}\left( a\right) \simeq a^{-\frac{3\lambda }{4+\lambda }}\left(
\ln a\right) ^{\frac{2+\lambda }{4+\lambda }}~,~\alpha =\ln a~,
\end{equation}%
and%
\begin{equation}
\sigma \left( a\right) =a^{\frac{12}{4+\lambda }-3}\left( \ln a\right) ^{-%
\frac{2}{4+\lambda }}\text{,}~~\alpha =\ln a.
\end{equation}

Finally, for $\lambda+4=0$ the closed-form solution for the physical
parameters are%
\begin{equation}
\alpha\left( t\right) =\frac{1}{6}\ln\left( -\frac{6\alpha_{1}}{%
V_{0}\cosh^{2}\left( 3\sqrt{\alpha_{1}}\left( t-t_{0}\right) \right) }%
\right) ~,
\end{equation}%
\begin{equation}
\beta\left( t\right) =\beta_{1}t+\beta_{0}~,
\end{equation}%
\begin{equation}
\phi\left( t\right) =\phi_{1}t+\phi_{0}\text{~},
\end{equation}
with constraint equation%
\begin{equation}
6\alpha_{1}-\frac{3}{2}\beta_{1}^{2}-\frac{\omega}{2}\phi_{1}^{2}=0.
\end{equation}

\subsection{Bianchi III \& Kantowski-Sachs spacetimes}

For $K\neq0$ and zero potential function, i.e. $V\left( \phi\right) =0$, we
consider the new variables $\phi=\Phi-\alpha$ and $\beta=B-4\Phi$, where
point-like Lagrangian (\ref{ff5}) becomes%
\begin{equation}
4e^{B}K-3\dot{\beta}^{2}-\left( \omega-12\right) \dot{\alpha}^{2}+2\left( 12%
\dot{B}+\omega\dot{\alpha}\right) \dot{\Phi}-\left( \omega+48\right) \dot{%
\Phi}^{2}=0.
\end{equation}

In the new variables the field equations read%
\begin{equation}
\ddot{\alpha}=\frac{2}{3}Ke^{B}~,~\ddot{\Phi}=\frac{2}{3\omega }\left(
\omega -12\right) Ke^{B}~,
\end{equation}%
\begin{equation}
\ddot{B}=\frac{2}{3}\left( \omega -16\right) Ke^{B}
\end{equation}%
with constraint equation%
\begin{equation}
4e^{B}K+3\dot{\beta}^{2}+\left( \omega -12\right) \dot{\alpha}^{2}-2\left( 12%
\dot{B}+\omega \dot{\alpha}\right) \dot{\Phi}+\left( \omega +48\right) \dot{%
\Phi}^{2}=0.
\end{equation}

Thus for $\left( \omega-16\right) \neq0$, the analytic solution for the
field equations is
\begin{equation}
B\left( t\right) =\ln\left( \frac{\omega B_{1}}{4\left( 16-\omega\right) }%
\cosh^{-2}\left( \frac{\sqrt{B_{1}}}{2}\left( t-t_{0}\right) \right) \right)
~,
\end{equation}%
\begin{equation}
\alpha\left( t\right) =-\frac{2\omega}{3\left( \omega-16\right) }\ln\left(
\cosh\left( \frac{\sqrt{B_{1}}}{2}\left( t-t_{0}\right) \right) \right)
+\alpha_{1}t+\alpha_{0}~,
\end{equation}%
\begin{equation}
\Phi\left( t\right) =-\frac{2\left( \omega-12\right) }{3\left(
\omega-16\right) }\ln\left( \cosh\left( \frac{\sqrt{B_{1}}}{2}\left(
t-t_{0}\right) \right) \right) +\Phi_{1}t+\Phi_{0}.
\end{equation}

On the other hand, for $\omega -16=0$, the analytic solution is%
\begin{equation}
B\left( t\right) =B_{1}t+B_{0}~,
\end{equation}%
\begin{equation}
\alpha \left( t\right) =\frac{2}{3B_{1}^{2}}e^{B_{0}+B_{1}t}K+\alpha
_{1}t+\alpha _{0}~,
\end{equation}%
\begin{equation}
\Phi \left( t\right) =\frac{1}{6B_{1}^{2}}e^{B_{0}+B_{1}t}K+\Phi
_{1}t+\alpha _{0}~.
\end{equation}%
Finally, the expansion rate has similar behaviour with the solution found
before.

According to our knowledge, these are the first anisotropic cosmological
analytic solutions in the literature in teleparallel dark energy theory.

\section{The Wheeler-DeWitt equation}

\label{sec6}

Consider the constraint Hamiltonian dynamical system of the form $H=$ $N%
\mathcal{H}$ with%
\begin{equation}
\mathcal{H=}\frac{1}{2}\mathcal{G}^{AB}p_{A}p_{B}+\mathcal{U}(\mathbf{{q})}%
\text{ and }\mathcal{H}\equiv 0~,
\end{equation}%
where $\mathcal{G}^{AB}$ is the minisuperspace, $p_{A}$ is the momentum
conjugate to the variable $q^{A}$ and $\mathcal{U}(\mathbf{{q})}$ is the
effective potential.

From the point-like Lagrangian (\ref{ch.07}) we find%
\begin{equation}
p_{\alpha }=\frac{12}{N}F\left( \phi \right) e^{3\alpha }\dot{\alpha}%
~,~p_{\beta }=-\frac{3}{N}F\left( \phi \right) e^{3\alpha }\dot{\beta}%
~,~p_{\phi }=-\frac{\omega }{N}F\left( \phi \right) e^{3\alpha }\dot{\phi}.
\end{equation}%
Hence, the Hamiltonian is written as%
\begin{equation}
H=N\left[ \frac{1}{2}\frac{e^{-3a}}{F\left( \phi \right) }\left( \frac{1}{12}%
p_{\alpha }^{2}-\frac{1}{3}p_{\beta }^{2}-\frac{1}{\omega }p_{\phi
}^{2}\right) -e^{3\alpha }F\left( \phi \right) \left( 2Ke^{-2\alpha +\beta
}+V\left( \phi \right) \right) \right] ,
\end{equation}%
that is, by using the constraint equation (\ref{ch.08}):
\begin{equation}
\mathcal{H=}\frac{1}{2}\frac{e^{-3a}}{F\left( \phi \right) }\left( \frac{1}{%
12}p_{\alpha }^{2}-\frac{1}{3}p_{\beta }^{2}-\frac{1}{\omega }p_{\phi
}^{2}\right) -e^{3\alpha }F\left( \phi \right) \left( 2Ke^{-2\alpha +\beta
}+V\left( \phi \right) \right) =0\,\text{.}
\end{equation}

The Wheeler-DeWitt equation follows under canonical quantization of the
constraint, that is, $\mathcal{H}\Psi(\mathbf{q})=0$, where $\Psi(\mathbf{q}%
) $ is the wavefunction of the quantized system.

For the Wheeler-DeWitt equation to remain invariant under conformal
transformations, the conformally Laplace operator is applied, such that $%
\mathcal{H}\Psi (\mathbf{q})=0\,\ $to be written as%
\begin{equation}
\mathcal{W}\left( \mathbf{q},\Psi \right) \equiv \left( \frac{1}{2}\Delta +%
\frac{n-2}{8(n-1)}\mathcal{R}-\mathcal{U}(\mathbf{q})\right) \Psi (\mathbf{q}%
)=0,  \label{wwd1}
\end{equation}%
in which $\mathcal{R}$is the Ricciscalar of the minisuperspace $\mathcal{G}%
_{AB}$ and $n=\dim \mathcal{G}$.

We consider the case $F\left( \phi \right) =e^{2\phi }$ and without loss of
generality, we set $N=e^{3\alpha +2\phi }$, thus, the Wheeler-DeWitt
equation becomes%
\begin{equation}
\mathcal{W}\left( \alpha ,\beta ,\phi ,\Psi \right) \equiv \left( \frac{1}{24%
}\frac{\partial ^{2}}{\partial \alpha ^{2}}-\frac{1}{6}\frac{\partial ^{2}}{%
\partial \beta ^{2}}-\frac{1}{2\omega }\frac{\partial ^{2}}{\partial \phi
^{2}}-e^{6\alpha +4\phi }\left( 2Ke^{-2\alpha +\beta }+V\left( \phi \right)
\right) \right) \Psi \left( \alpha ,\beta ,\phi \right) =0.  \label{wdw.01}
\end{equation}

\subsection{Quantum operators from symmetry vectors}

We briefly discuss the application of the theory of symmetries of
differential equations for the construction of quantum operators necessary
for the derivation of exact solutions for the Wheeler-DeWitt equation.

Assume now the generic vector field
\begin{equation}
\mathbf{Y=}\xi^{\mathbf{q}}\left( \mathbf{q},\Psi\right) \partial _{\mathbf{q%
}}+\eta\left( \mathbf{q},\Psi\right) \partial_{\Psi}
\end{equation}
defined in the jet space $\left\{ \mathbf{q},\Psi\right\} $, in which the
Wheeler-DeWitt equation (\ref{wwd1}) lies.

The vector field $\mathbf{Y}$ is the infinitesimal generator of the point
transformation between two points $P\left( \mathbf{q},\Psi \right)
\rightarrow P^{\prime }\left( \mathbf{q}^{\prime },\Psi ^{\prime }\right) $
\cite{Bluman,Olver}:%
\begin{equation}
\left( \mathbf{q}^{\prime },\Psi ^{\prime }\right) =\left( \mathbf{q},\Psi
\right) +\varepsilon \left( \xi ^{\mathbf{q}}\left( \mathbf{q},\Psi \right)
,\eta \left( \mathbf{q},\Psi \right) \right) ,
\end{equation}%
where $\varepsilon $ is an infinitesimal parameter such that $\varepsilon
^{2}\rightarrow 0$.

We say that the Wheeler-DeWitt equation (\ref{wwd1}) is invariant under the
point transformation with generator $\mathbf{X}$ if and only if%
\begin{equation}
\lim_{\varepsilon \rightarrow 0}\frac{\mathcal{W}^{\prime }\left( \mathbf{q}%
^{\prime },\Psi ^{\prime }\right) -\mathcal{W}\left( \mathbf{q},\Psi \right)
}{\varepsilon }=0.  \label{ftb.26}
\end{equation}%
When the latter condition is true, the vector field $\mathbf{Y}$ is a Lie
point symmetry for the differential equation.

\bigskip In \cite{AnIJ1}, it was found that the generic symmetry vector for
equation (\ref{wwd1}) is of the form

\begin{equation}
\mathbf{Y}=\xi\left( \mathbf{q}\right) \partial_{\mathbf{q}}+\left[ \frac{%
(2-n)}{2}\psi\left( \mathbf{q}\right) \Psi+\mu_{0}\Psi+\Omega\left( \mathbf{q%
}\right) \right] \partial_{\Psi}\mathbf{,}  \label{ftb.27}
\end{equation}
where $\xi\left( \mathbf{q}\right) $ is a conformal Killing vector field of
the minisuperspace $\mathcal{G}_{AB}$, with conformal factor $\psi\left(
\mathbf{q}\right) $, which satisfy the condition $\mathcal{L}_{\xi }\mathcal{%
U}(\mathbf{q})+2\psi\mathcal{U}(\mathbf{q})=0$, in which $\mathcal{L}_{\xi}$
is the Lie derivative with respect to the vector field $\xi$. $\Omega\left(
\mathbf{q}\right) $ satisfies the original equation and denotes the infinite
number of solutions of the original conformal Laplace equation.

For a given Lie symmetry $\mathbf{Y~}$of the conformal Laplace equation we
can determine in the normal variables the equivalent Lie-B\"{a}cklund vector
field $\mathbf{\hat{Y}}=\left( \frac{\partial \Psi }{\partial q^{J}}-\left(
\frac{2-n}{2}\psi +a_{0}\right) \Psi \right) \partial _{\Psi }$, which has
the property to transform a solution into a solution. That is, $\mathbf{\hat{%
Y}}\Psi =\mu _{1}\Psi $ from which it follows that%
\begin{equation}
\frac{\partial \Psi }{\partial q^{J}}-\left( \frac{2-n}{2}\psi +\alpha
_{0}\right) \Psi =\mu _{1}\Psi
\end{equation}%
which is nothing else than a quantum operator.

Recall that in the normal variables, the conservation laws of the classical
system generated by point symmetries can be written as the linear function
of the momentum $p_{J}$, that is $I_{J}=p_{J}$, from where it follows that
under canonical quantization
\begin{equation}
-i\frac{\partial\Psi}{\partial q^{J}}-I_{J}\Psi=0.
\end{equation}
From the later is clear that Noetherian conservation laws can be used to
determine differential invariants for the Wheeler-DeWitt equation and this
is an approach to relate the classical and quantum solutions.

\subsection{Exact solutions}

We assume now the case of Bianchi I universe, $K=0$, with the exponential
potential $V\left( \phi \right) =F_{0}e^{\lambda \phi }$. Then, with the use
of the Noether symmetry vectors we construct differential operators which
for the general case provide the analytic solution for the Wheeler-DeWitt
equation (\ref{wdw.01}):%
\begin{equation}
\Psi \left( \alpha ,\beta ,\phi \right) =\exp \left( c_{1}\beta -3\frac{c_{2}%
}{\left( 4+\lambda \right) ^{2}-3\omega }\left( 2c_{2}\lambda +8\alpha
+\omega \phi \right) \right) \left( \Psi _{1}J_{\kappa }\left( Z\right)
+\Psi _{2}Y_{\kappa }\left( Z\right) \right) ,
\end{equation}%
where $J_{\kappa }\left( \mathbf{q}\right) ,Y_{\kappa }\left( \mathbf{q}%
\right) $ are the Bessel functions of the first and second kind, while \ $%
\kappa =\frac{2}{3}\sqrt{\frac{\omega \left( 9c_{2}-c_{1}^{2}\left( \left(
4+\lambda \right) ^{2}-3\omega \right) \right) }{\left( \left( 4+\lambda
\right) ^{2}-3\right) ^{2}}},~Z=2\sqrt{\frac{2\omega V_{0}}{\left( 4+\lambda
\right) ^{2}-3\omega }}e^{3\alpha +\frac{1}{2}\phi \left( 4+\lambda \right)
} $, $\Psi _{1},~\Psi _{2}$, $c_{1}$ and $c_{2}$ are constants, while $c_{1}$
and $c_{2}$ are related with the constants of integration for the classical
solution.

Similarly, for the case with $K=0$ and $V\left( \phi \right) =0$, we
determine the exact solution%
\begin{equation}
\Psi \left( \alpha ,\beta ,\phi \right) =\exp \left( \frac{\omega \left(
c_{1}\left( 4\phi +\alpha +\beta \right) -3c_{2}\phi \right) +48\left(
c_{1}-c_{2}\right) \alpha -12c_{2}\beta }{3\left( 16-\omega \right) }\right)
\left( \Psi _{1}J_{\bar{\kappa}}\left( \bar{Z}\right) +\Psi _{2}Y_{\bar{%
\kappa}}\left( \bar{Z}\right) \right) ~,
\end{equation}%
in which $\bar{\kappa}=\frac{1}{3\left( \omega -16\right) }\sqrt{\omega
\left( \omega c_{1}^{2}+12\left( 2c_{1}-c_{2}\right) \left(
2c_{1}-3c_{2}\right) \right) }$ and $\bar{Z}=-2\sqrt{\frac{2K\omega }{%
16-\omega }}e^{2\left( \alpha +\phi \right) +\frac{\beta }{2}}$.

For specific values of the free parameters, the exact solutions have
different functional forms. However, they can easily be derived. Indeed, the
Wheeler-DeWitt equation with the use of the differential operators is
reduced to the Bessel second-order differential equation.

\section{Conclusions}

\label{sec7}

Anisotropic spacetimes describe the physical space in the early stages of
the universe and the determination of exact and analytic solutions is of
special interest, because the evolution of anisotropies in a given
gravitational theory can be studied analytically. In this study, we
investigated anisotropic cosmological solutions for the scalar-torsion
theory in the context of teleparallelism, where the physical space is
described by the locally rotational spaces of the Bianchi I, Bianchi III and
Kantowski-Sachs geometries.

For the scalar-torsion theory, we made use of a corresponding vierbein
fields defined in the proper frame for each background geometry such that
the limit of General Relativity to be recovered when the nonminimally
coupled scalar field becomes constant. The resulting field equations have
the property to have a point-like Lagrangian description. Furthermore, we
assume the existence of a potential function which drives the dynamics of
the scalar field and consequently of all of the physical parameters of the
theory.

We applied the Noether symmetry analysis in order to perform a
classification of the scalar field potential, by assuming the requirement
that there exist nontrivial variational point symmetries for the point-like
Lagrangian of the field equations. The later requirement is equivalent with
the existence of conservation laws for the field equations.\ That is ensured
by Noether's second theorem. From the classification scheme we are able to
identify specific functional forms for the scalar field potentials where the
resulting field equations form a Liouville integrable system. For that
potential functions we solved the field equations by using the additional
conservation laws and we wrote the resulting analytic solutions in
closed-form expressions.

It is known that there exist a unique relation between the Noether
symmetries of the classical field equations and of the Lie symmetries of the
Wheeler-DeWitt equation of quantum cosmology. Thus, we derived the
Wheeler-DeWitt equation and we determined differential operators which were
used to write the wavefunction in closed-form expression in terms of the
exponential and of the Bessel functions.

In a future study, we plan to investigate further the physical evolution of
the field equations for the scalar-torsion theory in an anisotropic
background geometry by applying other analytical techniques, such is the
analysis of the stationary points for the determination of asymptotic
solutions. The results of this analysis as that of the following study will
give important information for the initial value problem in scalar-torsion
cosmology.

\begin{acknowledgments}
The author thanks the Ionian University for the hospitality provided.
\end{acknowledgments}

\end{document}